# Optical Cages

V. Kumar*, J. P. Walker* and H. Grebel

The Electronic Imaging Center and the ECE department at NJIT, Newark, NJ 07102.
grebel@njit.edu

* Equally contributed

**Abstract:** We examine array of metal-mesh frameworks for their wide-band absorption. These take the form of quasi-crystal optical cages. An array of cages tends to focus the incoming radiation within each framework. An array of cage-within-cage funnels the radiation from the outer cage to its inner core even further.

**Introduction:** There is a growing interest in perfect absorbers [1], which do not transmit, nor reflect electromagnetic (EM) radiation. Metallo-dielectric absorbers on a sub-wavelength scale may broadly fall into several categories: resonating structures such as meta-materials [2-4], interferometric devices [5-7] and lossy thin film or surface guides [8-9], most of which are quasi-two dimensional. Here we propose a new class of quasi 3-dimensional hollow absorbers made of metal wire cages. Our simulations suggest that at the minimum, an array of quasi-crystal frameworks exhibits a large absorption coefficient (A~0.7) with a large bandwidth and are capable of trapping electromagnetic energy within them. A cage-within-cage framework exhibits a similar absorption coefficient (A~0.7) with much increased bandwidth.

First, let us introduce a metric by which we can compare all existing absorbers regardless of their frequency of operation. It is defined as the absorption-bandwidth product while normalizing it by the center frequency of operation. The bandwidth-to-frequency ratio is just the inverse of the quality factor Q, thus, the quality loss factor is simply, $L=A\Delta\lambda/\lambda=A/Q$. For example, an ultimate absorber, such as a frequency independent black body has an absorption coefficient of $A^{max}=1$, a very large bandwidth response and its center frequency is at the bandwidth center. In this case, the bandwidth is twice its center frequency: $Q^{max}=0.5$ and $L^{max}=2$. In the Yablonovitch limit [10], a frequency independent, mirror-clad, weakly absorbing film ($\alpha d=0.02$, n=3.5) in which the mirror suppresses the transmission, has it: $A=4n^2\alpha d/(1+4n^2\alpha d)=0.5$; Q~0.5; and L~1. Metal based resonators allow both transmission and reflection to occur, and are typically narrow band. For example [2,3], an excellent absorption coefficient of A~0.99 with Q~25 at microwave frequencies exhibits L~0.04. 'Thick' metal wire cage array (the wire thickness being 10% of the edge length) in air exhibits L~0.16 while a cage-within-cage exhibits L~0.28. 'Thin' metal wire cage-within-cage array (the wire thickness being 5% of the edge length) exhibits L~0.35

A Faraday Cage [11], a hollow structure made of knitted metal wires, shields its inner domain from external electromagnetic radiation through current loops at its surface. The openings in the wire mesh are very small compared with the effective radiation wavelength for the energy to be dissipated at the surface. Here we focus on cages whose dimensions and openings are of the order of the radiation wavelength. The excited dipoles are generating an internal field, which is as not totally frustrated as it happens for its homogeneous surface counterpart. The resonance wavelength is increasing upon varying the cage opening from being closed (a homogeneous icosahedron shell, SI section) to being large and defined by very thin metal wires.

Periodic metallo-dielectric structures (also known as screens, or metal meshes) have been studied in the past, and in particular in the long wavelength region – the wavelength region where the array pitch is of the order of, or smaller than the radiation wavelength [12-14]. This wavelength region is above the diffraction region. Stacked periodic metal screens resemble photonic crystals with a large index of refraction ratio [15]. Metal screens may be divided into two categories: inductive screens (metal films with a periodic array of holes which portray a transmission band) and capacitive screens (the complementary structure where metal structures are embedded in a dielectric and portray a reflection band). The screen's modes are made of local modes within each individual feature and extended propagating modes along the periodic array. At resonance, both local and extended modes form a composite standing wave. The screens exhibit negative index of refraction, or NIR (see SI section). For inductive screens, the NIR is exhibited throughout the wavelength band pass. For capacitive screens, the NIR region lies in the longer wavelength region beyond the reflection resonance. The roles of these screens is reversed when the screen's thickness becomes of the order of a wavelength. Despite large losses in the visible range, metals offer large index of refraction ratio when embedded in common dielectrics.

**Simulations:** In terms of a large absorption coefficient, array of cubic metal meshes (SI section) are not as efficient as icosahedrons. This suggests that there should be some degree of complexity to the framework. Yarn balls, made of azimuthally spaced rings around a common center are more effective than cubes but exhibit orientation dependency. Single elements are less efficient than an array frameworks, which suggests that coupling between adjacent elements is important. For simplicity, the icosahedrons were constructed of metal wires and were imbedded in air. The lattice constant for this square array was 1 micron and the wavelength range of interest was one octave between 1 to 2 microns. At shorter wavelengths, shorter than 1 micron, the array exhibits diffraction modes. Complex dielectric constant was selected [16]. The icosahedral edge and wire thickness varied. Periodic boundary conditions between the icosahedrons and perfect matching layers (PML) on the top and bottom of the computation cell were used. A CAD tool (Comsol) was employed in the analysis of the array. Results for icosahedral array each having an edge length of 0.4 microns and wire thickness of 0.04 microns are shown below. Optimizations with edge-length to wire-thickness ratio can be made and affect the resonance frequency.

Plots of the intensity coefficients for the transmission, T, reflection, R, and total absorption A (which is defined as, A=1-T-R) are provided in Fig. 1. The array of icosahedrons was illuminated by a plane wave, propagating along the z-direction with y-polarization.

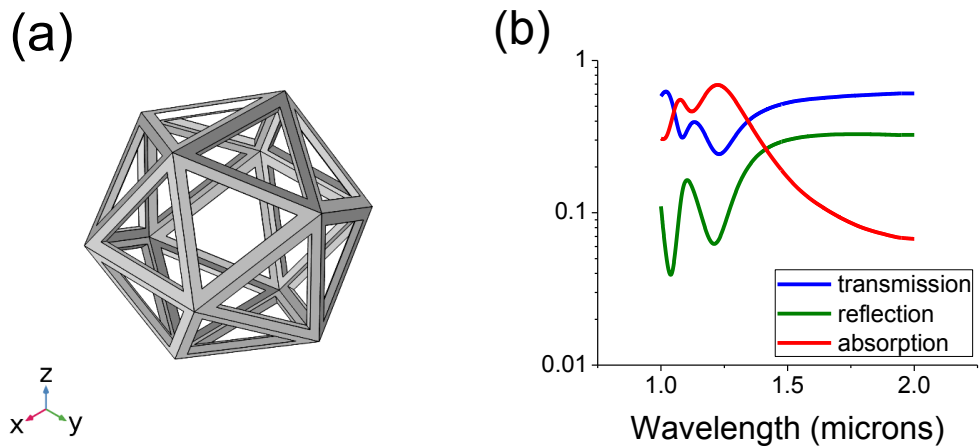

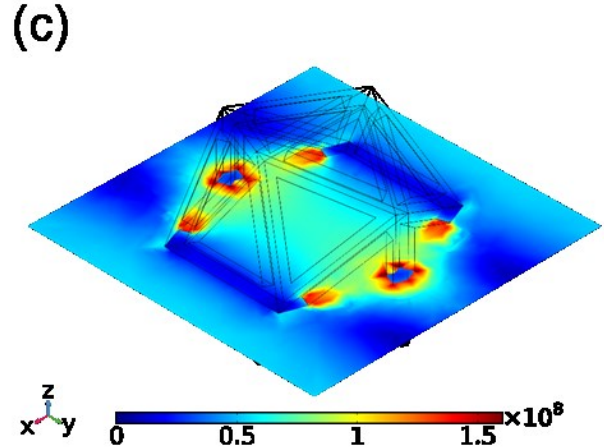

Fig. 1. (a) Copper-made icosahedral array: each edge measures 0.4 microns and the wire thickness was 0.04 microns. (b) Coefficients for the transmission intensity, T, reflection intensity, R, and absorption A (defined as, A=1-T-R) as a function of wavelength in microns. (c) Norm of the field distribution near each icosahedron's center. The wavelength is $\lambda$=1.23 microns and the light is polarized along the y-direction.

We identify two absorption peaks, at $\lambda$~1.1 and $\lambda$~1.22 microns, respectively which vary as a function of the icosahedron size. The absorption peaks are associated with the generation of shielding currents. Note the cold spots at the icosahedron corners. We calculated the dissipated electromagnetic intensity through shielding currents in the wires: $Q_h = \int dV \sigma\, \boldsymbol{E} \cdot \boldsymbol{E}$; with $\sigma$, the wire conductivity and $\boldsymbol{E}$, the driving local electric field. The dissipating loss constant is defined as, $P = Q_h/I_0$ where $I_0 \equiv 1$ W is the incident intensity. The difference between the A and P coefficients is practically zero in this wavelength range leading to the conclusion that the electric field has effectively excited resonating currents.

Silver is known to be a better conductor than copper. The peak absorption coefficient for a silver-made icosahedral array is smaller than for copper-made frameworks: the two peaks shifted to the shorter wavelengths with A~0.2. Similar analysis for a cage made of silicon results in P~0, which alludes to the fact that there are no dissipating currents above the bandgap.

An aligned, cage-within-cage framework is shown in Fig. 2. The enclosing (larger) icosahedron edge measures 0.4 microns and the enclosed (smaller) icosahedron edge measures 0.2 microns while the array pitch remains 1 micron. The wire thickness is 0.04 microns for the outer cage and 0.02 microns for the inner cage.

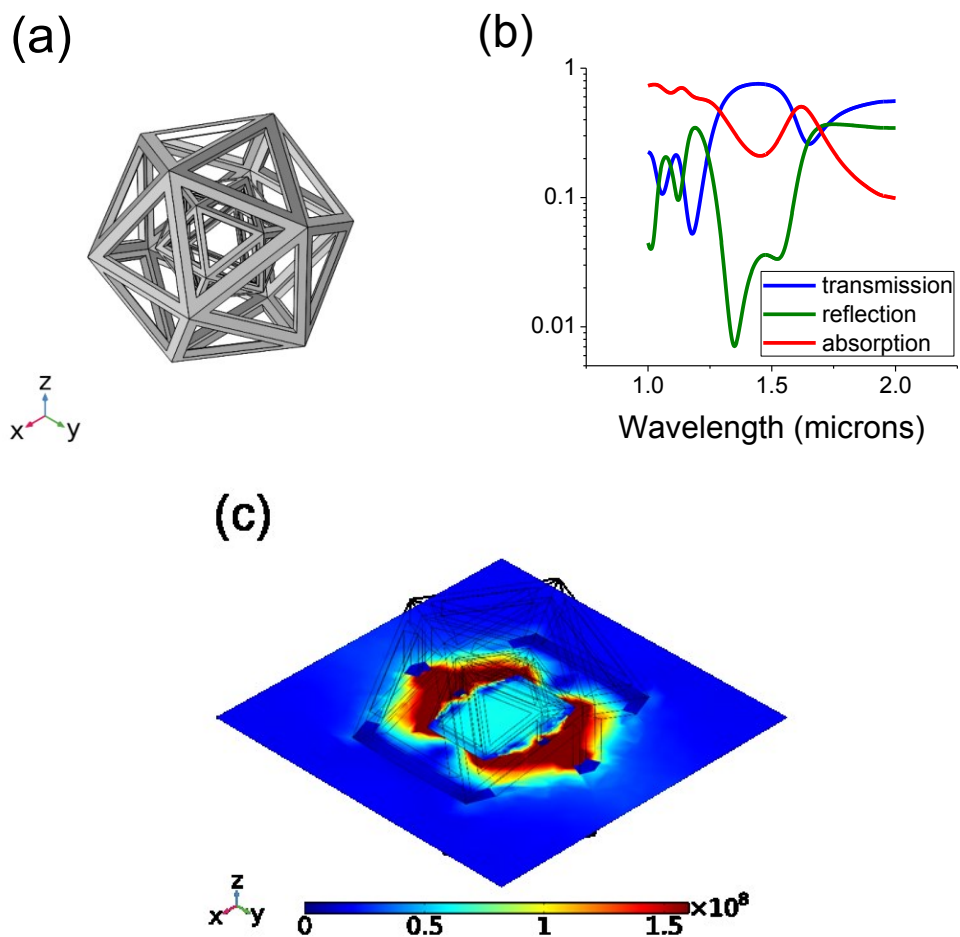

Fig. 2. (a) Aligned, copper-made, cage-within-cage. (b) Coefficients for the transmission intensity, T, reflection intensity, R, and absorption A (defined as, A=1-T-R) as a function of wavelength in microns. (c) Norm of the field distribution near the icosahedron's center at $\lambda$=1.59 microns (the longer wavelength peak). The incident light was polarized along the y-direction

The absorption intensity coefficient for a copper-made, cage-within-cage has not changed much, $A_{peak}$~0.75 at $\lambda$=1.15 microns; yet, the spectral width has increased and the spectra shifted to the shorter wavelengths. A third peak is seen at $\lambda$~1.6 microns and may be attributed to trapped radiation between the two icosahedra (Fig. 2c).

The peak absorption for a silver-made, cage-within-cage is much smaller than its copper-made counterpart: A~0.1 at $\lambda$=1.15 microns and A~0.35 at $\lambda$~1.6 microns.

At normal incidence, the absorption coefficient slightly depends on the polarization orientation. In Fig. 3 we show the absorption as a function of wavelength (in microns) for several azimuthal angles; these are defined between the x-axis and the incident linear polarization. The larger changes are observed at $\lambda$~1 microns whereas no change is observed at $\lambda$~1.6. The quality loss factor depends on the azimuthal angle; the thin-wire framework has larger values than the thick-wire one. In evaluating the quality loss factor we approximated the absorption with a Gaussian and assessed its width at half-maximum.

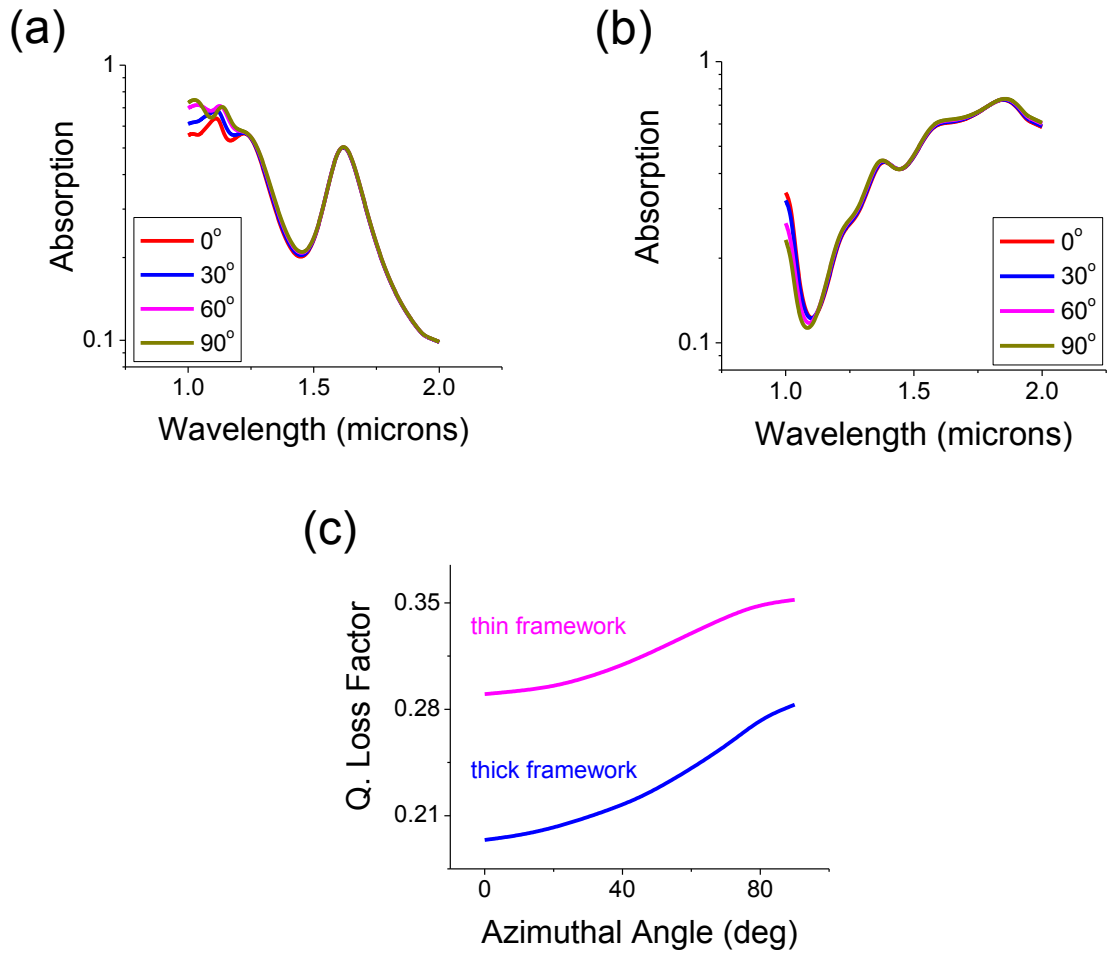

Fig. 3. A cage-within-cage. (a) Absorption intensity coefficient as a function of wavelength (in microns) for several azimuthal angle between the x-axis and the incident linear polarization. (b) A thin-wire framework. (c) Quality loss factor, L, as a function of azimuthal angle (in degrees) between the incident polarization and the x-axis.

Is icosahedral array the optimal cage structure? This is still an open question. Also puzzling is the smaller effect for silver-made with respect to copper-made frameworks; this could be attributed, in part to the lower plasmonic cut-off wavelength for silver.

Tuning the cage by means of acoustic waves [17-18] (similar to a wine-rack whose volume decreases upon extension), exploring these in the far-IR, embedding quantum dots inside them and study their nonlinear behavior are but a few exciting possibilities.

# Optical Cages

V. Kumar, J. P. Walker and H. Grebel

The Electronic Imaging Center and the ECE department at NJIT, Newark, NJ 07102.
grebel@njit.edu

Supplemental Information

**Capacitive screens:** At IR wavelengths, metals may be treated as perfect conductors. Transmission through a capacitive metal screen, each with feature made of a cross-shape is shown in Fig. S1a. The dip in transmission is indicative of a band reflection at the resonance frequency. The equivalent refractive index is shown in Fig. S1b. The array exhibits a NIR beyond the resonance wavelength.

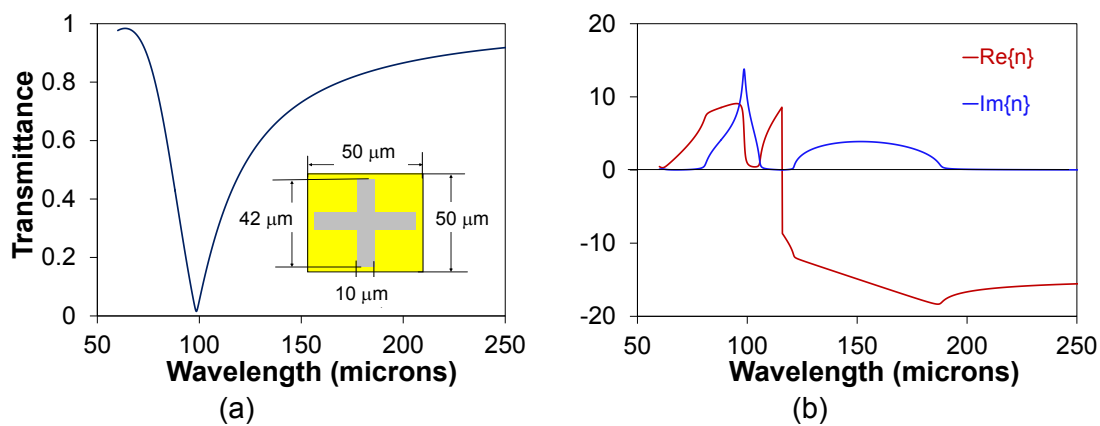

Fig. S1. (a) An example of transmittance through a capacitive grid made of a metal cross on a dielectric at far-IR and (b) its related refractive index (D. Moeller and H. Grebel, unpublished)

An array made of thin shell icosahedrons is shown in Fig. S2. The absorption is not as large as for the wire counterpart and little electric field fills the structure.

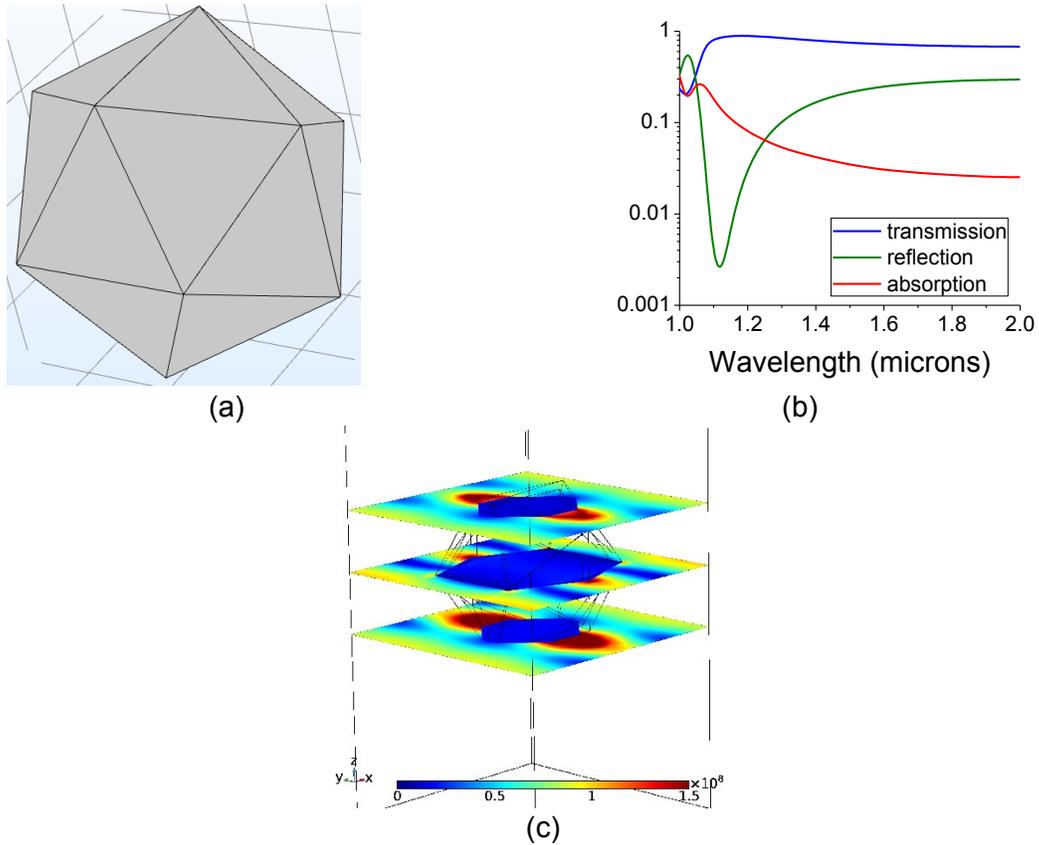

Fig. S2. Thin (t=0.02 microns) hollow icosahedral metal framework. (a) The facet configuration for an edge length of 0.4 microns and a lattice constant of 1 microns. (b) Intensity coefficient for transmission, T, reflection, R, and absorption A (A=1-T-R). (c) Norm of the field intensity cross-sections at $\lambda$=1.0714 microns – the maximum absorption wavelength.

A framework of cubic metal wires is shown in Fig. S3. Its edge is 0.9 microns (almost overfilling its pitch of 1 microns) and the thickness of the wires is 0.045 microns. Note that the icosahedrons are quite close to one another. The efficiency of cubic cage array is limited; larger absorption may be achieved upon an increase of its wire thickness.

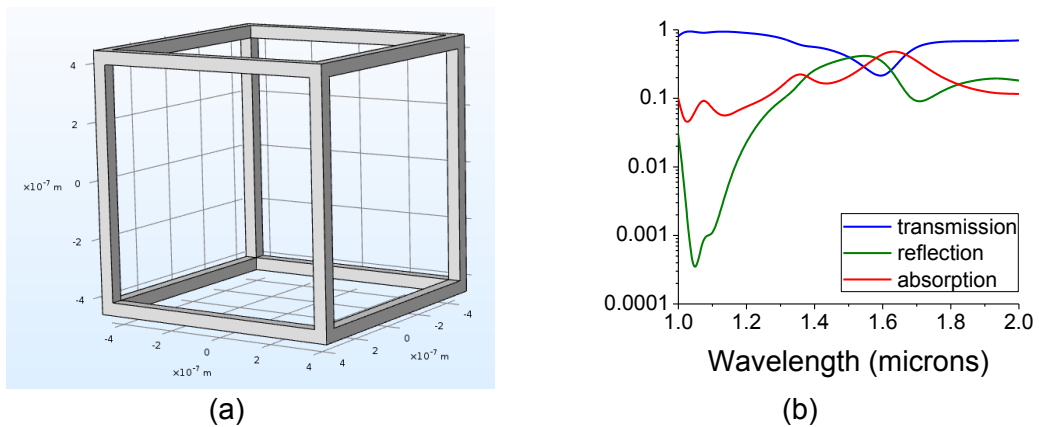

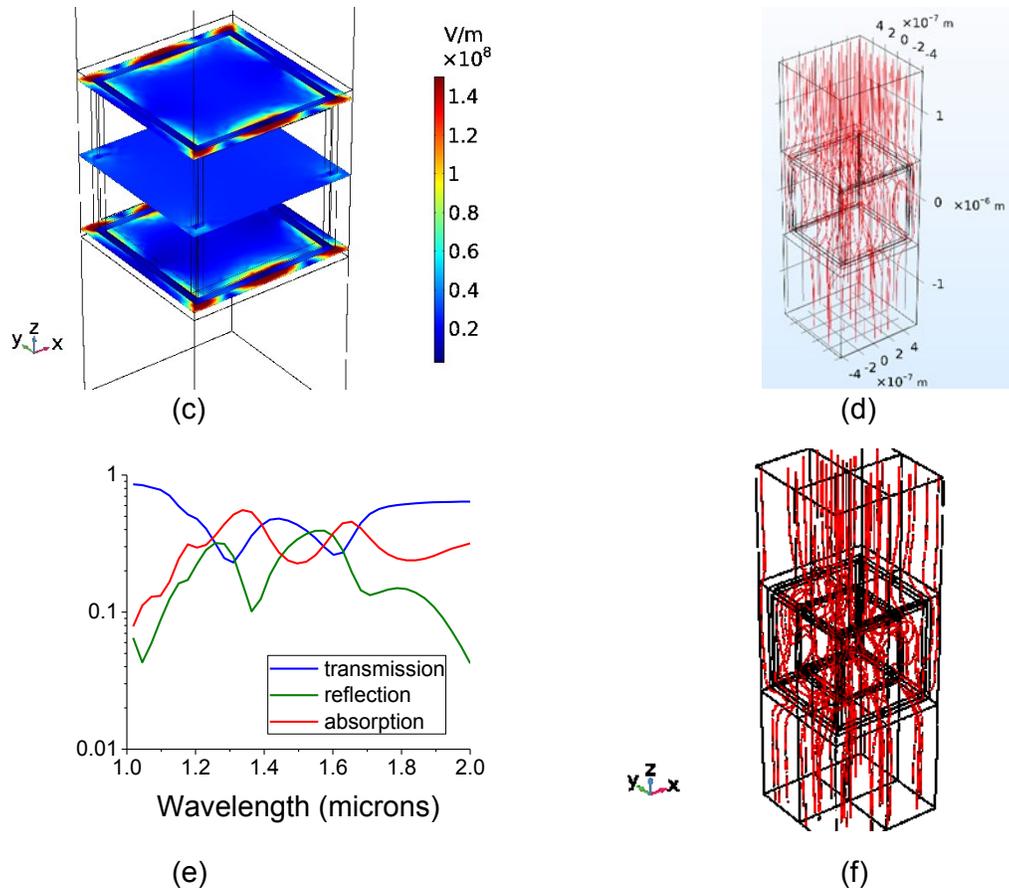

Fig. S3. Cubic metal framework. (a) The wire configuration for an edge length of 0.9 microns and a lattice constant of 1 microns. (b) Intensity coefficient for transmission, T, reflection, R, and absorption A (A=1-T-R). The peak absorption at λ=1.63 microns is ca 0.4. (c) Norm of the field intensity cross-sections. The plane wave illuminates the structure from the top and the polarization is along the y-direction. (d) The distribution of intensity lines at λ=1.63 microns. (e) Intensity coefficients for a cube-within-cube and (f) the distribution of the intensity lines at λ=1.63 microns. The structure further focuses the incident radiation.